\documentclass[reprint,amsmath,amssymb]{revtex4-2}

\usepackage{xcolor}
\usepackage[utf8]{inputenc}
\usepackage{physics}
\usepackage{amsfonts}
\usepackage{dcolumn}
\usepackage{bm}
\usepackage{wrapfig}
\usepackage[normalem]{ulem}
\usepackage[colorlinks=true, hyperindex, breaklinks, linkcolor=blue, urlcolor=blue, citecolor=blue]{hyperref}
\usepackage{setspace}
\usepackage{lineno}
\usepackage[T1]{fontenc}
\usepackage{svg}
\usepackage{stfloats}
\usepackage{booktabs}
\usepackage{tabularx}
\usepackage{comment}
\usepackage{xcolor}
\usepackage[normalem]{ulem}

\usepackage{bibunits}

\defaultbibliography{references}
\defaultbibliographystyle{naturemag}

\usepackage{graphicx}
\usepackage{float}
\usepackage{amssymb, amsthm, amsmath}
\usepackage{accents}
\usepackage{mathptmx} 
\usepackage{textcomp}

\usepackage{cleveref}
\crefname{equation}{equation}{equations}
\Crefname{equation}{equation}{equations}
\crefname{table}{Table}{Tables}
\Crefname{table}{Table}{Tables}
\crefname{figure}{Fig.}{Figs.}
\Crefname{figure}{Fig.}{Figs.}
\crefname{section}{Section}{sections}
\Crefname{section}{Section}{Sections}
\crefname{subsection}{Section}{Sections}
\Crefname{subsection}{Section}{Sections}

\usepackage[caption=false]{subfig}

\newcommand{\phantomsubfloat}[1]{
	{
		\captionsetup[subfigure]{labelformat=empty}
		\subfloat[][]{#1}
	}%
}

\begin{document}

\title{\fontsize{11.25}{14}\selectfont Qubit Noise Sensing via Induced Photon Loss in a Superconducting Cavity\\}

\author{Nitzan Kahn}
\thanks{These authors contributed equally to this work.}
\author{Dror Garti}
\thanks{These authors contributed equally to this work.}
\author{Uri Goldblatt}
\author{Lalit M. Joshi}
\author{Fabien Lafont}
\author{Serge Rosenblum}
\affiliation{
\vspace{3pt}Department of Condensed Matter Physics, Weizmann Institute of Science, Rehovot, Israel
}
\graphicspath{{plots/}}

\begin{abstract}
\noindent 
Characterizing noise in superconducting qubits is essential for improving coherence and gate performance. Conventional noise-sensing methods typically use the qubit itself as the sensor, which limits both accessible bandwidth and applicability during driven operation. Here, we measure qubit frequency noise by detecting the photon loss it induces in a coupled high-Q superconducting cavity. We use repeated mid-circuit qubit measurements with post-selection to separate this induced loss from intrinsic cavity decay. We validate the protocol using injected noise and show that the extracted loss scales as expected with the applied noise strength. We place an upper bound of $0.70\times10^3\,\mathrm{rad}^2/\,\mathrm{s}$ on the intrinsic qubit frequency-noise power spectral density at the cavity–qubit detuning of 508 MHz. The protocol opens access to a higher-frequency spectral window than standard qubit-based spectroscopy and may enable noise characterization during strong driving.

\end{abstract}

\maketitle

Superconducting qubits couple strongly to their electromagnetic environment, making them susceptible to noise sources that induce energy relaxation, dephasing, and gate errors. These sources include voltage and current fluctuations in control circuitry, quasiparticle tunneling \cite{catelani2011relaxation}, and electric and magnetic field fluctuations associated with microscopic two-level system (TLS) defects \cite{muller2019towards}. 
Identifying and characterizing such noise processes is central to advancing qubit performance.

    Standard noise detection methods use the qubit itself as a probe~\cite{Sung2019Non-GaussianSensor, suter2016colloquium, souza2011robust, alvarez2011measuring, Najera-Santos2023High-sensitivityQubit}, which limits their accessible frequency range: Ramsey interferometry is sensitive at sub-kHz frequencies~\cite{Yan2012SpectroscopyQubit}, dynamical decoupling reaches up to $20~\rm{MHz}$~\cite{Bylander2011NoiseQubit}, and driven-evolution protocols, including spin locking and strongly driven Rabi spectroscopy, up to $\sim300~\rm{MHz}$~\cite{Yan2013Rotating-frameEvolution, yan2018distinguishing, Sung2021Multi-levelSpectroscopy, Yoshihara_PhysRevB.89.020503}. 
    Microwave single-photon detectors~\cite{Inomata2016SingleSystem, Lescanne2020IrreversiblePhotons} and relaxation spectroscopy \cite{Bylander2011NoiseQubit, Astafiev2004QuantumQubit} provide access to the GHz band, leaving a window of several hundred MHz without a dedicated probe.  
    This window is especially relevant during strongly driven operations, such as readout and parametric gates, where control tones can produce high-frequency noise~\cite{Slichter2012Measurement-inducedNoise}.

Superconducting cavities offer a complementary detection platform \cite{reagor2013reaching, reagor2016quantum, krasnok2024superconducting}. In particular, they have proven effective for precision sensing \cite{Checchin2022MeasurementResonator, Read2023PrecisionSensitivity}, notably in 
axion and dark photon dark matter searches, where they serve as resonant detectors for weak microwave signals \cite{alesini2019galactic, Braggio2025Quantum-EnhancedCounter, tang2024first, Romanenko2023SearchCavities, agrawal2024stimulated, berlin2022searches, DixitDarkMatter}.
Recent advances \cite{Takenaka2025Three-DimensionalLifetime, Romanenko2020Three-DimensionalS,kudra2020high} have enabled these cavities to achieve millisecond-scale photon lifetimes when integrated with chip-based superconducting qubits \cite{Chakram2021Seamless, Kim2025UltracoherentControl}, making them natural candidates for characterization of qubit noise.
 
\begin{figure}[b!]
 \vspace{-10pt} 
 \centering 
 \includegraphics[scale=0.56]{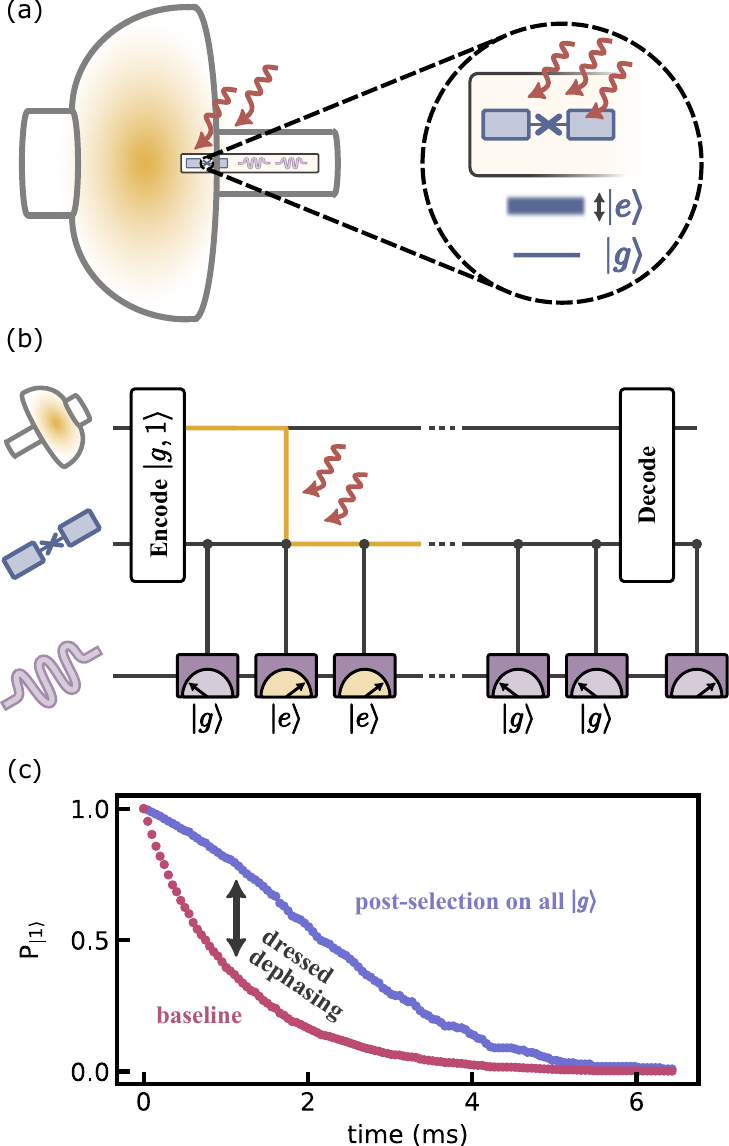}
 \phantomsubfloat{\label{fig1:scheme a}}
	\phantomsubfloat{\label{fig1:scheme b}}
	\phantomsubfloat{\label{fig1:scheme c}}
	
 \vspace{-15pt} 
 \caption{Detecting qubit frequency noise with a superconducting cavity. (a) Experimental setup. A half-elliptical cavity is coupled to a transmon qubit subject to frequency noise (red arrows), corresponding to fluctuations in the qubit's resonance frequency. (b) The sensing protocol. A single photon is initialized in the cavity, and the qubit is measured at fixed intervals $T_{\rm{m}}$ during the cavity evolution. The evolution time $t=NT_{\rm{m}}$ is scanned by varying the number of measurements $N$. In a dressed-dephasing event (red arrows), the cavity excitation is transferred to the qubit and detected by a subsequent measurement. Finally, the cavity population is mapped back onto the qubit and read out. (c) Simulated protocol output. The baseline cavity decay curve (pink) decays exponentially at the total loss rate $\kappa$. The non-exponential post-selected curve (purple), conditioned on all $\ket{g}$ outcomes, decays more slowly because detected photon loss events have been discarded. The dressed-dephasing rate $\kappa_\mathrm{dd}$ is extracted by comparing both curves.}
\end{figure} 

In this work, we use a high-quality-factor (high-$Q$) cavity to probe frequency noise in a chip-based superconducting qubit (Fig. \ref{fig1:scheme a}). Our approach uses dressed dephasing~\cite{Note_dressed_dephasing}, in which qubit frequency noise drives energy exchange between the cavity and the qubit, opening an additional photon loss channel whose rate is proportional to the frequency-noise spectral density at the cavity-qubit detuning ~\cite{boissonneault2009dispersive,Pietikainen2024StrategiesElectrodynamics}. Previous work characterized this effect by driving a low-$Q$ resonator and observing qubit heating~\cite{Slichter2012Measurement-inducedNoise}. Here, we instead store a single photon in a millisecond-lifetime cavity and infer qubit frequency noise from the resulting photon loss. Repeated mid-circuit qubit measurements with post-selection separate qubit-noise-induced photon loss from intrinsic cavity decay, allowing the dressed-dephasing rate $\kappa_{\rm{dd}}$ to be extracted. We first validate the method by injecting controlled frequency noise and measuring the cavity lifetime response as a function of noise amplitude. At sufficiently low injected-noise amplitudes, no dressed-dephasing signal is resolvable, constraining the intrinsic dressed-dephasing rate to $\kappa_{\mathrm{dd}}<(0.42 \, \textrm{s})^{-1}$ and the corresponding frequency-noise power spectral density (PSD) at the cavity-qubit detuning $\Delta/2\pi = 508 \, \mathrm{MHz}$ to $S_{\delta\omega}(\Delta)<0.70\times10^3\, \mathrm{rad}^2/ \,\mathrm{s}$. This approach extends noise spectroscopy into the high-frequency window inaccessible to standard qubit-based methods. Because the sensor is an undriven cavity mode rather than the qubit itself, the method remains applicable while the qubit is driven.

\begin{figure}[h]
 \centering 
 \includegraphics[]
 {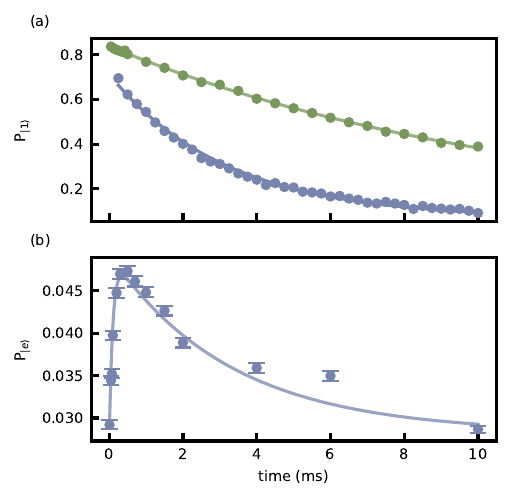}
 \phantomsubfloat{\label{fig2:scheme a}}
	\phantomsubfloat{\label{fig2:scheme b}}
 \vspace{-3.8\baselineskip}
 \vspace{10pt} 
 \caption{Characterizing the impact of dressed dephasing on the cavity-qubit system. (a) The cavity single-photon lifetime without injected noise (green), $T_1^\mathrm{c} = 11.3~\textrm{ms}$, and with injected noise (blue), $T_1^\mathrm{c} = 3.2~\textrm{ms}$. Solid lines are exponential fits. Error bars are smaller than the markers. (b) Qubit excited-state probability as a function of time in the presence of injected noise. Dressed dephasing transfers the cavity excitation to the qubit, resulting in a temporary increase in the qubit's excited-state population. Here, unlike the full sensing protocol (Fig.~\ref{fig3}), the qubit population is measured only once at the end of each evolution time, without mid-circuit measurements, to directly observe the excitation transfer.
 The solid line is a simulation with $\kappa_\textrm{dd}$ as the sole free parameter, yielding $\kappa_\mathrm{dd} = (4.6~\mathrm{ms})^{-1}$. All other parameters are fixed from independent measurements (Table~\ref{tab:system_parameters_full}). Error bars represent $\pm1\sigma$ standard error.}
 \label{fig2}
\end{figure}

Dressed dephasing arises when frequency noise at the cavity-qubit detuning 
drives photon exchange between the qubit and the cavity. When the qubit is initialized in the ground state and the cavity is populated by a single photon, the dressed-dephasing-induced photon loss rate is given by \cite{boissonneault2009dispersive}
\begin{align}
\label{eq1}
 \kappa_{\mathrm{dd}} = 4\frac{g^2}{\Delta^2}S_{\delta\omega}(\Delta),
\end{align}
where $g$ is the cavity-qubit coupling strength and $S_{\delta\omega}(\Delta)$ is the frequency noise PSD
evaluated at the cavity-qubit detuning frequency $\Delta\gg g$ (Supplemental Material~\ref{app:dd_transduction}~\cite{SM_dd_rate}). 
Equation~\eqref{eq1} shows that the transduction of qubit frequency noise into cavity photon loss is suppressed by the small hybridization $g^2/\Delta^2$ of the cavity mode with the qubit.
Since $\kappa_{\mathrm{dd}}$ is directly proportional to $S_{\delta\omega}(\Delta)$, isolating $\kappa_{\mathrm{dd}}$
provides a direct measure of the noise at this frequency.
However, $\kappa_\mathrm{dd}$ enters the total cavity decay rate $\kappa = \kappa_\mathrm{dd} + \kappa_0$ alongside the intrinsic loss rate $\kappa_0$, so a standard lifetime measurement alone cannot distinguish noise-induced loss from intrinsic cavity decay. 

To separate the two contributions, we perform repeated mid-circuit measurements of the qubit during the cavity evolution and compare two decay curves constructed from the same data (Figs.~\ref{fig1:scheme b} and \ref{fig1:scheme c}). Disregarding all qubit measurement outcomes yields the unconditional photon survival probability $P_{\ket{1}}(t) = e^{-\kappa t}$, which reflects the total loss rate. Post-selecting on trajectories in which every qubit measurement returns the ground state discards runs in which a cavity excitation was transferred to the qubit via a dressed-dephasing event and detected by a subsequent measurement. The remaining trajectories comprise two cases: those in which the photon survived, and those in which it was lost without detection — either through intrinsic cavity decay or through a dressed-dephasing event followed by qubit relaxation before the next measurement. The conditional photon survival probability is therefore 

\begin{equation}
P_{\ket{1}}^g(t) = \frac{e^{-\kappa t}}{e^{-\kappa t}+(1 - \xi) (1-e^{-\kappa t})},
\label{eq2}
\end{equation}
where the detection probability $\xi$ is the fraction of photon loss events producing a detectable qubit excitation within a single measurement interval $T_\mathrm{m}$:
\begin{equation}
\xi = \frac{\kappa_\mathrm{dd}}{\Gamma - \kappa} \cdot \frac{e^{-\kappa T_\mathrm{m}} - e^{-\Gamma T_\mathrm{m}}}{1 - e^{-\kappa T_\mathrm{m}}},
\label{eq3}
\end{equation}
with $\Gamma$ the qubit decay rate.
In the limit $T_\mathrm{m}\ll 1/\Gamma$, nearly every dressed-dephasing event is detected before the qubit relaxes ($\xi\to\kappa_\mathrm{dd}/\kappa$), while for $T_\mathrm{m}\gg 1/\Gamma$, qubit relaxation between measurements erases the signal ($\xi\to 0$). A derivation of Eqs. (\ref{eq2})-(\ref{eq3}) is provided in Supplemental Material \ref{app:post-selection_derivation}~\cite{SM_post_selection}. 

To realize the cavity-based sensing scheme, 
we dispersively couple a transmon qubit (resonance frequency $\omega_\textrm{q}/2\pi = 3.800~\textrm{GHz}$, relaxation time $T_1^\textrm{q}\equiv1/\Gamma = 71~\mu \mathrm{s}$) to a high-$Q$ 3D cavity \cite{Milul2023SuperconductingTime,Romanenko2020Three-DimensionalS} (resonance frequency $\omega_\mathrm{c}/2\pi = 4.308~\textrm{GHz}$, single-photon lifetime $T_1^\textrm{c} = 11.3~\textrm{ms}$). The qubit is also dispersively coupled to a Purcell-filtered readout resonator connected to a quantum-limited amplification chain for high-fidelity single-shot readout (Supplemental Material \ref{app:device_specification}~\cite{SM_device} and refs.~\cite{heidler2021non, oriani2025niobium}).

\begin{figure}[b]
 \centering 
 \includegraphics[]{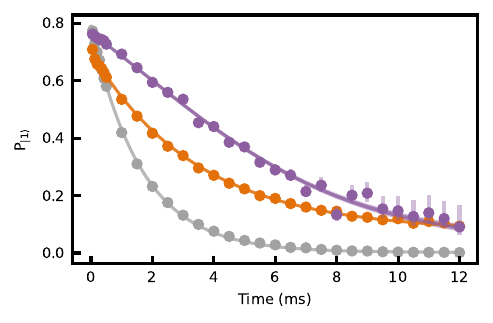}
 \vspace{-2\baselineskip}
 \vspace{10pt} 
 \caption{Unconditional (orange) $P_{\ket{1}}(t)$ and post-selected (purple) $P_{\ket{1}}^g(t)$ single-photon survival probabilities in the presence of injected frequency noise (at a different noise amplitude than in Fig.~\ref{fig2}). Solid lines are fits to Eq. (\ref{eq2}) (Supplemental Material section \ref{app:bayesian_statistics}~\cite{SM_bayesian} and Ref.~\cite{pymc2023}) with shaded regions (narrower than the linewidth) denoting the 68\% highest density interval. Error bars are $1\sigma$ confidence intervals. From the fit we extract $\kappa_{\mathrm{dd}} = (5.2\pm0.1~\mathrm{ms})^{-1}$, corresponding to an induced noise PSD of $S_{\delta\omega}(\Delta) = 57 \times 10^3~\mathrm{rad}^2~/~\mathrm{s}$, and an unconditional decay rate of
 $\kappa =(3.46 \pm 0.02~\mathrm{ms})^{-1}$. 
 The fraction of experimental shots surviving post-selection (gray) decays exponentially at a rate of $(1.64 \pm 0.01~\mathrm{ms})^{-1}$, primarily due to thermal qubit excitations and false-positive measurements. Each data point corresponds to 20{,}000 experimental repetitions.
 }
 \label{fig3}
\end{figure}

To characterize the impact of dressed dephasing on the cavity, we inject controlled frequency noise into the qubit and measure the resulting change in cavity lifetime. We prepare a single photon in the cavity using a parametric sideband interaction \cite{pechal2014microwave, rosenblum2018cnot} (Supplemental Material \ref{app:protocol}~\cite{SM_protocol}) and drive the qubit with two noise-broadened microwave tones with amplitudes $a_1(t)$ and $a_2(t)$. The beat note of the tones produces an oscillating AC Stark shift through the four-wave mixing Hamiltonian $\hat{H}_\textrm{noise}\propto \, a_1(t)a_2^*(t)\hat\sigma_z +\textrm{h.c.}$, generating qubit frequency noise at their difference frequency (Supplemental Material \ref{app:noise_derivation}~\cite{SM_noise_derivation} and \ref{app:noise_engineering}~\cite{SM_noise_engineering}). Scanning this difference frequency, we observe a reduction in the single-photon lifetime caused by dressed dephasing (Supplemental Material Fig.~\ref{fig:pg_vs_beat_multiple}). When the difference frequency equals $\Delta$, the cavity lifetime reduces from 11.3 ms to 3.2 ms (Fig. \ref{fig2}a).

We confirm that this increased photon loss originates from dressed dephasing by monitoring the qubit excited-state population~\cite{Jin2015ThermalQubit} rather than the photon survival probability (Fig. \ref{fig2}b). The excited-state population initially rises, reflecting excitation transfer from the cavity to the qubit, and subsequently relaxes to its steady-state value as the added excitation dissipates. This steady-state value of  2.9\% is slightly above the thermal population of  2.3\% measured without injected noise (Table \ref{tab:system_parameters_full}), reflecting a small additional heating by the noise tones. A simulation of the system dynamics with $\kappa_\mathrm{dd}$ as the sole free parameter reproduces both the lifetime reduction and the transient qubit heating (Supplemental Material \ref{app:simulations}~\cite{SM_simulations} and Ref. \cite{johansson2012qutip}), indicating that dressed dephasing accounts for the observed photon loss.

\begin{figure}[t] 
 \centering 
 \includegraphics[]{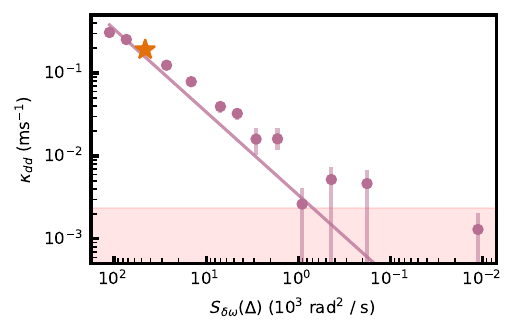}
 \vspace{-2\baselineskip}
 \vspace{10pt} 
 \caption{
 Extracted dressed-dephasing rate $\kappa_\mathrm{dd}$ versus the injected noise PSD, $S_{\delta\omega}(\Delta)$.
 The solid line (pink) shows the prediction of Eq.~(\ref{eq1}), using independently calibrated injected noise powers and evaluated with $g/2\pi=14.8~\rm{MHz}$ (Supplemental Material~\ref{app:g_extraction}~\cite{SM_g_extraction}). The shaded region (red) indicates the 68\% credible upper limit on $\kappa_\mathrm{dd}$, determined from the three lowest-power data points, each consisting of 20,000 repetitions, at which the injected noise is predicted to contribute below this limit. This limit is set by the finite statistical precision of the combined post-selected lifetime estimate. Error bars denote 68\% Bayesian highest density intervals, and the star (orange) indicates the noise PSD corresponding to the data in Fig.~\ref{fig3}.
 }
 \label{fig4}
\end{figure}

Next, we benchmark the full sensing protocol using injected noise. We prepare a single cavity photon, perform repeated mid-circuit measurements of the qubit every $T_\mathrm{m}=7.45~\mathrm{\mu}$s, 
and post-select on trajectories in which every measurement yields $\ket{g}$~\cite{Goldblatt2024RecoveringFeedback}. The mid-circuit measurements have no measurable effect on $\kappa_0$ (Fig.~\ref{fig:measurement_rate_effect}). Fitting both the unconditional and post-selected photon survival probabilities (Fig. \ref{fig3}) to Eq. (\ref{eq2}), we extract $\kappa_\mathrm{dd} = (5.2\pm0.1~\mathrm{ms})^{-1}$, in agreement with the injected noise strength independently calibrated via the measured AC Stark shift (Supplemental Material~\ref{app:noise_engineering}~\cite{SM_noise_engineering}). The extracted intrinsic loss rate $\kappa_0=(10.2^{+0.5}_{-0.4}~\mathrm{ms})^{-1}$ indicates that the protocol correctly separates
noise-induced loss from intrinsic cavity decay. The corresponding
detection probability is $\xi = 0.63$: since the measurement interval
satisfies $T_\mathrm{m} \ll 1/\Gamma$, nearly every dressed-dephasing event is
detected before the qubit relaxes, and $\xi$ is limited primarily by the fraction of loss events due to dressed dephasing
($\kappa_\mathrm{dd}/\kappa$).

We further validate the protocol by repeating it across a range of injected noise amplitudes (Fig.~\ref{fig4}). At high noise power, the extracted $\kappa_{\mathrm{dd}}$ scales linearly with the noise PSD, as expected from Eq.~(\ref{eq1}). 

Finally, at sufficiently low noise power, we can no longer resolve $\kappa_\textrm{dd}$, and the extracted values become upper bounds set by our statistical resolution.
Using these values, we bound the dressed-dephasing rate to $\kappa_{\rm{dd}}<(0.42~\rm{s})^{-1}$ and the qubit's intrinsic frequency-noise PSD at $\Delta/2\pi=508~\rm{MHz}$ to $S_{\delta\omega}(\Delta)<0.70\times10^3~\mathrm{rad}^2/\rm{s}$. This bound is evaluated using $N=60,000$ repetitions, and further averaging would lower the floor as $N^{-1/2}$. 
The extracted bound places the frequency noise in our fixed-frequency transmon two orders of magnitude below that inferred in a flux-sensitive qubit~\cite{Slichter2012Measurement-inducedNoise}.

In summary, we demonstrated a cavity-based technique for sensing qubit frequency noise through photon loss induced by dressed dephasing. A key feature of the protocol is repeated mid-circuit qubit readout with post-selection, which separates the noise-induced loss channel from intrinsic cavity decay. Applying the protocol to a transmon qubit coupled to a 3D superconducting cavity, we place an upper bound on the intrinsic dressed-dephasing rate of $(0.42 \, \textrm{s})^{-1}$ at $508\,\textrm{MHz}$, corresponding to $S_{\delta\omega}(\Delta)<0.70\times10^3\,\mathrm{rad}^2/\mathrm{s}$ at 60,000 repetitions. The protocol opens access to a high-frequency spectral window that is difficult to probe with conventional qubit-based methods and is directly relevant to strongly driven qubit operation. 

    An extension of the proposed protocol to broadband sensing could be realized by using a high-$Q$ multimode flute cavity~\cite{Chakram2021Seamless} to sample $S_{\delta\omega}$ at a dense set of mode frequencies spanning $\sim 2~\rm{GHz}$. 
    Alternatively, the detuning $\Delta$ could be adjusted in situ by tuning either the sensor or the qubit under study. Piezoelectric tuning of the cavity resonance~\cite{c2016piezoelectric} moves the sampling frequency while leaving the qubit untouched, whereas the qubit itself could be tuned with a microwave-induced Stark shift or by using a flux-tunable transmon~\cite{koch2007charge,chavez2022weakly}. In the latter case, the added qubit flux noise would increase the measured frequency-noise signal.

Improved sensitivity may be achieved through longer cavity lifetimes, stronger cavity–qubit hybridization, or larger cavity
photon number~\cite{Slichter2012Measurement-inducedNoise}, the latter requiring a dispersive shift large enough to resolve individual photon numbers. Sensitivity can also be gained without changing the hardware: rather than discarding every trajectory containing an excited-state outcome, an analysis of the full measurement record, for instance a hidden Markov model of the detected excitations, would extract additional information about individual dressed-dephasing events while retaining shots that post-selection currently rejects, reducing the number of repetitions needed to reach a given noise floor.

Beyond probing qubit frequency noise, this method could help constrain background estimates for cavity-based dark-matter searches~\cite{Braggio2025Quantum-EnhancedCounter,tang2024first,Romanenko2023SearchCavities}. Qubit-induced dressed dephasing removes signal photons from the cavity and, in reverse, converts qubit excitations into spurious cavity photons that mimic candidate events. Both processes are governed by $\kappa_{\rm{dd}}$, which the protocol isolates from the intrinsic cavity decay, turning this qubit-induced background into a directly measurable quantity.

\vspace{12pt}
\textit{Acknowledgments---} We thank Yuval Waserman for discussions on the statistical analysis presented in this work. 
This research was sponsored by the Army Research Office and was accomplished under Grant Number W911NF-25-1-0196. S.R. is the incumbent of the Rabbi Dr Roger Herst Career Development Chair.

\bibliography{references}

\clearpage
\onecolumngrid

\setcounter{section}{0}
\setcounter{subsection}{0}
\setcounter{equation}{0}
\setcounter{figure}{0}
\setcounter{table}{0}

\renewcommand{\thesection}{S\arabic{section}}
\renewcommand{\theequation}{S\arabic{equation}}
\renewcommand{\thefigure}{S\arabic{figure}}
\renewcommand{\thetable}{S\arabic{table}}
\renewcommand{\theHsection}{S\arabic{section}}
\renewcommand{\theHsubsection}{S\arabic{section}.\arabic{subsection}}
\renewcommand{\theHequation}{S\arabic{equation}}
\renewcommand{\theHfigure}{S\arabic{figure}}
\renewcommand{\theHtable}{S\arabic{table}}
\crefname{figure}{Fig.}{Figs.}
\Crefname{figure}{Fig.}{Figs.}
\crefname{table}{Table}{Tables}
\Crefname{table}{Table}{Tables}
\crefname{equation}{Eq.}{Eqs.}
\Crefname{equation}{Eq.}{Eqs.}
\crefname{section}{Section}{Sections}
\Crefname{section}{Section}{Sections}


\begin{center}
{\Large\bfseries Qubit Noise Sensing via Induced Photon Loss in a Superconducting Cavity\par}
\vspace{1em}
{\large Supplemental Material\par}
\vspace{1em}

Nitzan Kahn,$^{*}$ Dror Garti,$^{*}$ Uri Goldblatt, Lalit M. Joshi, Fabien Lafont, and Serge Rosenblum\par
\vspace{0.5em}
\textit{Department of Condensed Matter Physics, Weizmann Institute of Science, Rehovot, Israel}\par
\vspace{0.5em}
$^{*}$These authors contributed equally to this work.
\end{center}

\vspace{1em}
\label{app:supplemental_Material}
\twocolumngrid

\section{Device specification}
\label{app:device_specification}

The experimental device comprises three main elements: a superconducting cavity, a transmon qubit, and a readout resonator. The cavity is a separate realization of the half-elliptical design reported in Ref.~\cite{Milul2023SuperconductingTime}, fabricated from two high-purity niobium components that are electron-beam welded to eliminate seam losses. Its half-elliptical geometry is employed to concentrate the electric field at the cavity center, enabling strong coupling to the qubit while minimizing photon loss due to interactions with lossy surfaces~\cite{heidler2021non}. 
 To support long electromagnetic field lifetimes~\cite{Romanenko2020Three-DimensionalS, Milul2023SuperconductingTime}, the cavity undergoes initial chemical etching to remove machining damage and surface contamination. Periodic water-based etching is applied to mitigate the degrading effects of oxygen exposure~\cite{oriani2025niobium}.

\begin{figure}[h]
 \centering
 \includegraphics[width=0.8\columnwidth]{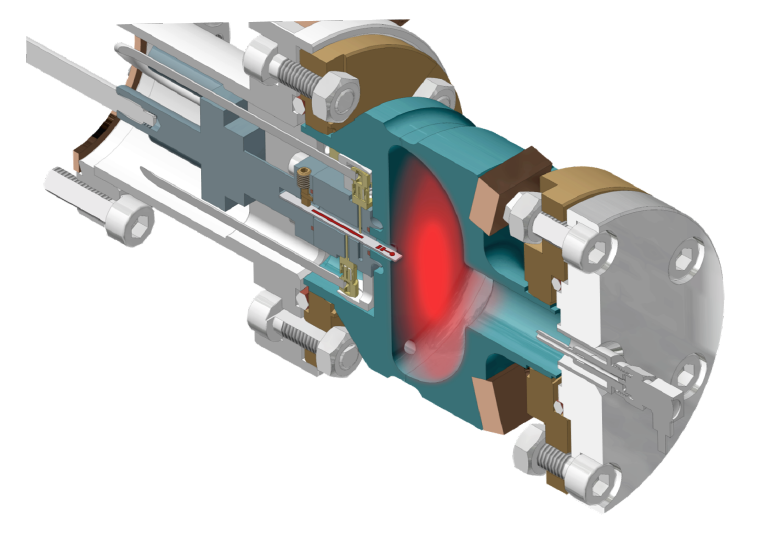}
 \vspace{-10pt}
 \caption{Sectional view of the experimental device, showing the 
half-elliptical 3D niobium cavity and the transmon chip protruding 
into it via a coaxial waveguide.}
 \label{fig:cavity_sectional_view}
\end{figure}

The aluminum transmon qubit and readout resonator are fabricated on a sapphire chip using electron-beam evaporation. The chip is housed in a narrow coaxial waveguide and protrudes approximately 1 mm into the cavity. It is accessed via input and output couplers integrated into the waveguide. The entire device is enclosed within an aluminum shield lined with copper sheets (Figs. \ref{fig:cavity_sectional_view} and~\ref{fig:fridge_inner_diagram}).

\begin{table*}[]
 \centering
 \begin{tabular}{l l l}
  \toprule
  \textbf{Parameter} & \textbf{Description} & \textbf{Value} \\
  \midrule
  $\omega_\mathrm{c}/2\pi$ & Cavity resonance frequency    & 4.308 GHz \\
  $T_1^\mathrm{c}$   & Cavity single-photon lifetime    & $11.3 \pm\ 0.2$ ms \\
  $T_2^\mathrm{c}$   & Cavity coherence time      & $3.4\pm\ 0.2$ ms \\
  $\chi/2\pi$  & Cavity-qubit dispersive shift   & $170\pm 20$ kHz (*) \\
  $g/2\pi$   & Cavity-qubit coupling     & $14.8 \pm 1$ MHz \\
  \midrule
  $\omega_\mathrm{q}/2\pi$ & Transmon resonance frequency    & 3.800 GHz \\
  $K_\mathrm{q}/2\pi$  & Transmon anharmonicity      & 124 MHz \\
  $T_1^\mathrm{q}$    & Transmon lifetime       & $71.0 \pm\ 1.0 \mu $s \\
  $T_2^\mathrm{q}$    & Transmon coherence time     & $35.0 \pm 2.0 ~\mu$s \\
  $T_{2E}^\mathrm{q}$   & Transmon Hahn-echo coherence time   & $42.0 \pm 1.0~\mu$s \\
  $\bar{n}_{\mathrm{th}}$ & Transmon average thermal population & $2.3 \pm 0.1 \%$ \\
  $\chi_\mathrm{qr}/2\pi$ & Transmon-readout dispersive shift   & $1.05 \pm 0.01$\ MHz \\
  \midrule
  $\omega_\mathrm{r}/2\pi$ & Readout resonance frequency    & 7.82 GHz \\
  $T_\mathrm{r}$    & Readout lifetime       & $607.0 \pm 8.0$ ns \\
  \bottomrule
 \end{tabular}
 \caption{
  System parameters and their respective values. Values obtained through calculation, rather than direct measurement, are indicated with an asterisk.
 }
 \label{tab:system_parameters_full}
\end{table*}

\section{Hamiltonian and Parameters}
\label{app:hamiltonian_and_parameters}

We can express the free evolution of the system using the following Hamiltonian:

\begin{equation}
\label{eq:System_Hamiltonian}
\begin{split}
\hat{H}/\hbar &= \omega_\mathrm{r} \hat{r}^{\dagger}\hat{r} + 
\omega_\mathrm{q} \hat{q}^{\dagger}\hat{q} + 
\omega_\mathrm{c} \hat{c}^{\dagger}\hat{c} \\
&\quad - \frac{K_\mathrm{r}}{2}\hat{r}^{\dagger}\hat{r}^{\dagger}\hat{r}\hat{r} -
\frac{K_\mathrm{q}}{2}\hat{q}^{\dagger}\hat{q}^{\dagger}\hat{q}\hat{q} -
\frac{K_\mathrm{c}}{2}\hat{c}^{\dagger}\hat{c}^{\dagger}\hat{c}\hat{c} \\
&\quad - \chi\hat{c}^{\dagger}\hat{c}\hat{q}^{\dagger}\hat{q} -
\chi_{qr}\hat{r}^{\dagger}\hat{r}\hat{q}^{\dagger}\hat{q} -
\chi_{cr}\hat{r}^{\dagger}\hat{r}\hat{c}^{\dagger}\hat{c}\,,
\end{split}
\end{equation}

where $\hat{r}, \hat{q}, \hat{c}$ are the annihilation operators of the readout, transmon qubit, and cavity modes, respectively. The descriptions and values of the parameters are listed in Table \ref{tab:system_parameters_full}.

    \section{Derivation of the dressed-dephasing photon-loss rate}
\label{app:dd_rate}

\subsection{Effective Longitudinal Noise Derivation}
\label{app:noise_derivation}
We derive the effective longitudinal frequency-noise Hamiltonian $\propto a_1(t)a_2^*(t)\hat\sigma_z$  generated by injecting two off-resonant microwave tones with amplitudes $a_1(t)=\varepsilon_1(t)\,e^{-i\omega_1 t}$ and $a_2(t)=\varepsilon_2(t)\,e^{-i\omega_2 t}$, where $\omega_j$ and $\varepsilon_j(t)$ are
 the center frequency and envelope of tone $j$, respectively. We start from the driven transmon Hamiltonian
\begin{equation}
\hat H_q/\hbar = \omega_q\hat{q}^\dagger \hat{q} -\frac{E_J}{\hbar}\left[\cos\hat{\varphi}-1+\frac{\hat{\varphi}^2}{2}\right]+ \sum_{j=1,2}\left[\varepsilon_j(t)\,e^{-i\omega_j t}\,\hat q^\dagger
+ {\rm h.c.}\right],
\end{equation}

\noindent where $E_J$ is the Josephson energy and $\hat\varphi= \varphi_{\rm zpf} ( \hat q+\hat q^\dagger)$ with $\varphi_{\rm zpf}$ the zero-point phase fluctuation across the junction.\\

\noindent Moving to the displaced frame
$\hat D(\beta)=\exp[\beta\hat q^\dagger-\beta^*\hat q]$ and choosing $\beta(t)$
to cancel the terms linear in $\hat q$, i.e.\
$\dot\beta=-i\omega_q\beta-i\sum_j\varepsilon_j e^{-i\omega_j t}$, gives in
steady state to leading order
\begin{equation}
\label{eq:beta_ss}
\beta(t) = \beta_1(t)e^{-i\omega_1 t} + \beta_2(t)e^{-i\omega_2 t},
\qquad
\beta_j(t) = \frac{\varepsilon_j(t)}{\omega_j-\omega_q}.
\end{equation}
What remains is the undriven transmon with a classically displaced junction
phase $\hat\varphi = \varphi_{\rm zpf} \left[ \hat q+\hat q^\dagger+\beta(t)+\beta^*(t) \right]$
Assuming the displacement is weak (i.e., $\varphi_{\rm zpf}|\beta(t)+\beta^*(t)|\ll1$), we can expand the Josephson cosine to fourth order, yielding
\begin{equation}
-E_J\left[ \cos\hat\varphi-1+\frac{\hat\varphi^2}{2} \right]\simeq - \frac{E_J}{24}\hat\varphi^4=-\frac{E_J\varphi_{\rm zpf}^4}{24}
\left[
\hat X+B(t)
\right]^4,
\end{equation}
\noindent where we defined $\hat X=\hat q+\hat q^\dagger$ and $B(t)=\beta(t)+\beta^*(t)$.

\noindent We retain the quartic contribution containing two powers of the classical displacement and two powers of the qubit operator $B^2(t)\hat X^2$. Using $[\hat q,\hat q^\dagger]=1$, the number-conserving part of $\hat X^2$ is
\begin{equation}
\hat X^2 = (\hat q+\hat q^\dagger)^2 = 
\hat q^{\dagger 2} + \hat q^2+2\hat q^\dagger\hat q+1
\rightarrow 2\hat q^\dagger\hat q+1 .
\end{equation}

\noindent The next step is the rotating-wave approximation: in the interaction picture, $\hat q^2$ and $\hat q^{\dagger 2}$ rotate at $\mp2\omega_q$ and average out. Dropping the identity contribution, we obtain

\begin{equation}
\hat H_{\rm Stark} = -\frac{E_J\varphi_{\rm zpf}^4}{2}
B^2(t)\hat q^\dagger\hat q .
\end{equation}

\noindent Writing the transmon anharmonicity as
\begin{equation}
K_q = \frac{E_J\varphi_{\rm zpf}^4}{2\hbar},
\end{equation}

\noindent we obtain
\begin{equation}
\frac{\hat H_{\rm Stark}}{\hbar} = -K_q B^2(t)\hat q^\dagger\hat q .
\end{equation}

\noindent We now extract the part of $B^2(t)$ that is slow compared with the microwave carrier frequencies. Since

\begin{equation}
B^2(t) = \beta^2(t)+\beta^{*2}(t)+2|\beta(t)|^2,
\end{equation}

\noindent the terms $\beta^2(t)$ and $\beta^{*2}(t)$ oscillate near $2\omega_1$, $2\omega_2$, and $\omega_1+\omega_2$, and are neglected. Thus,

\begin{align}
B^2(t) \rightarrow 2|\beta(t)|^2
&= 2 \Bigl[ |\beta_1(t)|^2 + |\beta_2(t)|^2 \\
&\quad + \beta_1(t)\beta_2^*(t)e^{-i\omega_b t}
+ \beta_1^*(t)\beta_2(t)e^{i\omega_b t} \Bigr]
\nonumber
\end{align}

\noindent where $\omega_b=\omega_1-\omega_2$. The first two terms result in static AC Stark shifts
\begin{equation}
\nu_i=-2K_q|\beta_i(t)|^2.  
\label{eq:stark}
\end{equation}
Keeping only the non-static beat-note terms gives
\begin{equation}
\frac{\hat H_{\rm Stark}(t)}{\hbar} \rightarrow -2K_q
\left[
\beta_1(t)\beta_2^*(t)e^{-i\omega_b t}
+
\beta_1^*(t)\beta_2(t)e^{i\omega_b t}
\right]
\hat q^\dagger\hat q .
\end{equation}

\noindent Projecting onto the two-level computational subspace, $\hat q^\dagger\hat q\rightarrow(1+\hat\sigma_z)/2$, and dropping the identity term, the beat-note Stark shift becomes a longitudinal qubit-frequency modulation,
\begin{equation}
\hat H_{\rm noise}(t) = \hbar\delta\omega(t)\hat\sigma_z ,
\end{equation}

\noindent with
\begin{equation}
\label{eq:delta_omega_t}
\delta\omega(t) = -K_q
\left[
\beta_1(t)\beta_2^*(t)e^{-i\omega_b t}
+
\beta_1^*(t)\beta_2(t)e^{i\omega_b t}
\right].
\end{equation}
$\delta\omega(t)$ is defined as half the fluctuation of the qubit transition frequency, following the notation of Ref.~\cite{boissonneault2009dispersive}.
\noindent Finally, identifying the drive amplitude with the phase displacement, i.e., $a_j(t) \propto \beta_j(t)e^{-i\omega_j t}$, we retrieve the AC Stark shift from the main manuscript:
\begin{equation}
\label{eq:H_noise}
\hat H_{\rm noise}(t) \propto a_1(t)a_2^*(t)\hat\sigma_z + {\rm h.c.}
\end{equation}

\noindent This shows that the two-tone four-wave-mixing process produces a
longitudinal modulation of the qubit transition frequency centered at the difference
frequency $\omega_b=\omega_1-\omega_2$. 

\subsection{Conversion of Qubit Frequency Noise Into Cavity Photon Loss}
\label{app:dd_transduction}
We next introduce hybridization of the transmon qubit with the cavity mode to show the conversion of the longitudinal modulation $\hbar \delta\omega(t)\hat\sigma_z$ into cavity photon loss. The derivation below applies to any such modulation regardless of its origin, and therefore to the intrinsic qubit frequency noise as well as to the engineered noise of Sec.~\ref{app:noise_derivation}.

Following Refs.~\cite{boissonneault2009dispersive, Pietikainen2024StrategiesElectrodynamics},
we move to the dispersive frame via
$\hat U = \exp[\lambda(\hat\sigma_+\hat c - \hat\sigma_-\hat c^\dagger)]$, where $\lambda \equiv g/\Delta$ with
$g$ the qubit-cavity coupling and $\Delta \equiv \omega_{\rm{c}}-\omega_{\rm{q}}$ the
detuning ($|\Delta| \gg g$). The noise operator
becomes
\begin{equation}
  \hat U^\dagger \hat H_{\rm noise} \hat U = \hbar\delta\omega(t)\hat{\sigma}_z
  + 2\hbar\lambda\,\delta\omega(t)\big(\hat\sigma_+\hat c + \hat\sigma_-\hat c^\dagger\big)
  + \mathcal{O}(\lambda^2),
  \label{eq:sz_dressed}
\end{equation}
adding a cavity-qubit beam-splitting interaction of amplitude $2\lambda\,\delta\omega(t)$ to
the residual longitudinal term.

In the interaction picture, the beam-splitter operators rotate at the detuning
$\hat\sigma_+\hat c \to \hat\sigma_+\hat c\,e^{-i\Delta t}$ and
$\hat\sigma_-\hat c^\dagger \to \hat\sigma_-\hat c^\dagger\,e^{+i\Delta t}$. 
When $\delta\omega(t)$ has a $\Delta$ frequency component, it 
cancels the rotation of the
$|g,1\rangle\!\leftrightarrow\!|e,0\rangle$ term, which then converts
a cavity photon into a qubit excitation.
 
From first-order perturbation theory, the
amplitude to populate $|e,0\rangle$ from $|g,1\rangle$ after a time $t$ is
\begin{equation}
  c_{e0}(t) = -\,2i\lambda\int_0^t \delta\omega(t')\,e^{-i\Delta t'}\,dt'.
  \label{eq:amp}
\end{equation}
Using $\langle e,0|\hat\sigma_+\hat c|g,1\rangle = 1$, and averaging its modulus
squared over the noise gives
\begin{equation}
  \big\langle |c_{e0}(t)|^2\big\rangle
  = 4\lambda^2\!\int_0^t\!\!\int_0^t\! dt'\,dt''\;
    \big\langle \delta\omega(t')\,\delta\omega(t'')\big\rangle\,
    e^{-i\Delta(t'-t'')}.
  \label{eq:prob_double}
\end{equation}

Stationarity makes the autocorrelation depend only on the time difference,
$\langle\delta\omega(t')\delta\omega(t'')\rangle = C(\tau)$, $\tau=t'-t''$, whose Fourier
transform is the power spectral density,
\begin{equation}
  S_{\delta\omega}(\omega) = \int_{-\infty}^{\infty} d\tau\;C(\tau)\,e^{+i\omega\tau}.
  \label{eq:PSD_def}
\end{equation}
In the Born--Markov limit,
where $t$ exceeds the correlation time of $\delta\omega(t)$, the double integral
reduces to 
\begin{equation}
    \langle |c_{e0}(t)|^2\rangle \simeq 4\lambda^2\, S_{\delta\omega}(-\Delta)\, t
= 4\lambda^2\, S_{\delta\omega}(\Delta)\, t,
\end{equation}
the last step using that
$S_{\delta\omega}$ is even for classical noise. 
Its slope is the dressed-dephasing photon-loss rate,
\begin{equation}
  \kappa_{\mathrm{dd}} = (2\lambda)^2\,S_{\delta\omega}(\Delta)
  = 4\,\frac{g^2}{\Delta^2}\,S_{\delta\omega}(\Delta),
  \label{eq:kappa_dd_deriv}
\end{equation}
i.e.\ Eq.~\eqref{eq1} of the main text, with the hybridization $g^2/\Delta^2$
setting the transduction of qubit frequency noise into cavity loss. The
conjugate term $\hat\sigma_-\hat c^\dagger e^{+i\Delta t}$ drives the reverse
process $|e,0\rangle\to|g,1\rangle$ at the same rate
$4\lambda^2 S_{\delta\omega}(\Delta)=\kappa_{\mathrm{dd}}$, so the forward and reverse dressed-dephasing collapse operators (Sec.~\ref{app:simulations}) carry equal weight. 
The protocol of the main text initializes the state $|g,1\rangle$, where $\bar n = 1 \ll n_{\rm crit} = \Delta^2/4g^2$, so the experiment operates in the linear dispersive regime, where the photon-number-dependent corrections present in the general master equation of Ref.~\cite{boissonneault2009dispersive} are negligible and $\kappa_{\rm{dd}}$ retains the form of Eq.~(\ref{eq:kappa_dd_deriv}).

\begin{figure}[h]
 \centering 
 \includegraphics[]{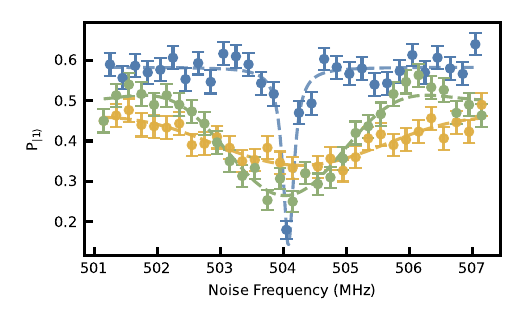}
 \vspace{-3.3\baselineskip}
 \vspace{10pt} 
\caption{Cavity photon survival probability $P_{\ket{1}}$ as a function of the 
applied noise frequency, measured without phase randomization (blue) 
and with phase randomization every 500~ns (orange) and 200~ns (green). The main text uses phase randomization every 300~ns. 
The dashed lines are fits to Lorentzian (blue) or $\mathrm{sinc}^2$ (green, orange) profiles. Without randomization, the dip 
is narrow and centered at the cavity-qubit detuning 
$\Delta = \omega_\mathrm{c} - \omega_\mathrm{q}$, consistent 
with dressed dephasing. 
Phase randomization broadens the noise spectrum, producing a wider and shallower dip. 
The detuning here is $\Delta/2\pi = 504$~MHz; the main-text data were 
acquired at $\Delta/ 2\pi = 508$~MHz due to a system drift.}
\label{fig:pg_vs_beat_multiple}
\end{figure}

\section{Frequency noise injection}
\label{app:noise_engineering}

To validate our sensing protocol, we engineer controlled frequency noise on the qubit by applying two microwave tones. The tones, with amplitudes $a_1(t)$ and $a_2(t)$ centered at frequencies  $\omega_1/2\pi= 6.000~\rm{GHz}$ and $\omega_2/2\pi = 5.493~\rm{GHz}$, produce via four-wave mixing a qubit AC Stark shift $\delta\omega(t)\hat\sigma_z$ oscillating at their difference frequency $\omega_b = \omega_1 - \omega_2$, with 
\begin{equation}
\delta\omega(t) = -K_q\left(\tilde \beta_1(t)\,\tilde \beta_2^*(t)+\textrm{c.c.}\right),
\end{equation}
and $\tilde \beta_i(t) = \beta_i(t) e^{-i\omega_i t + i\phi_i(t)}$ (Sec.~\ref{app:noise_derivation}). To make the frequency fluctuation stochastic, the phase $\phi_2(t)$ is drawn uniformly at random every interval $T$, while the first tone remains coherent. Because independent phase blocks are uncorrelated, the autocorrelation of $\delta\omega(t)$ vanishes for time separations exceeding~$T$ and decays linearly to zero within a single block. The corresponding frequency-noise power spectral density is
\begin{align}
&S_{\delta\omega}(\omega) = \nonumber\\
&K_q^2
|\beta_1|^2\,|\beta_2|^2\, T \,\left[
\mathrm{sinc}^2\!\left( \frac{(\omega - \omega_b)\,T}{2} \right)+ \mathrm{sinc}^2\!\left( \frac{(\omega + \omega_b)\,T}{2} \right)\right],
\end{align}
with peaks centered at $\pm\omega_b$ and bandwidth $2\pi/T$ set by the phase-randomization rate.

The injected noise strength can be independently calibrated using the Stark shift produced by each tone. When applied separately, each tone shifts the qubit frequency by $\nu_i  =-2K_q |\beta_i|^2$ (Eq.~\eqref{eq:stark}). The peak PSD can therefore be written as
\begin{equation}
S_{\delta\omega}(\omega=\omega_b) = \frac{1}{4} |\nu_1|\,|\nu_2|\, T\cdot D,
\end{equation}
where the duty cycle $D=0.53$ arises from the presence of qubit measurements, during which no noise is injected. For the data in Fig.~\ref{fig3}, the individually measured Stark shifts are $\nu_1/2\pi = -93$~kHz and $\nu_2/2\pi = -314$~kHz. With our chosen phase randomization time $T = 300\;\mathrm{ns}$, this yields $S_{\delta\omega}(\Delta) = 46 \times 10^3\,\mathrm{rad}^2/\mathrm{s}$, consistent with the value of $57 \times 10^3\,\mathrm{rad}^2/\mathrm{s}$ extracted from the sensing protocol. We attribute the difference to systematic uncertainty in the absolute transduction factor $4g^2/\Delta^2$, dominated by the calibration of $g$ (Sec. \ref{app:g_extraction}), which enters quadratically. This is an overall scale factor and does not affect the linear scaling of $\kappa_\mathrm{dd}$ with injected noise power.

\begin{figure}[t]
 \centering 
 \includegraphics[]{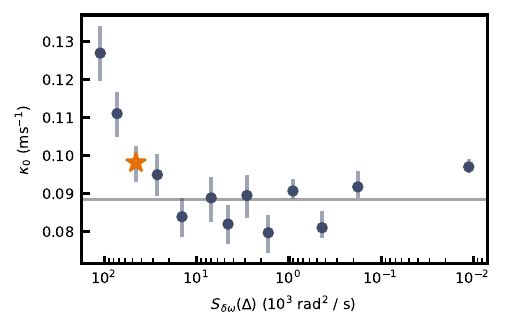}
 \vspace{-2\baselineskip}
 \vspace{10pt} 
 \caption{ Intrinsic cavity decay rate $\kappa_0$ as a function of injected noise PSD. The values for $\kappa_0$ are extracted from the unconditional and post-selected photon survival probabilities using Bayesian inference (Supplemental Material~\ref{app:bayesian_statistics}). At high noise amplitudes, we observe an increase in $\kappa_0$ compared to the baseline value in the absence of injected noise (gray line). The noise amplitudes used in Figs.~\ref{fig2} and~\ref{fig3} (orange star marks the latter) were chosen to remain in the regime where $\kappa_0$ is not significantly affected by the injected noise.
 }
\label{fig:kappa_0}
\end{figure}

We calibrate $\omega_b$ to match the cavity--qubit detuning by scanning $|\omega_1 - \omega_2|$ and measuring the cavity single-photon survival probability at each value (Fig.~\ref{fig:pg_vs_beat_multiple}); the detuning is identified as the frequency of maximum photon loss. At high noise amplitudes, we observe a reduction in the intrinsic cavity lifetime (Fig.~\ref{fig:kappa_0}), whose origin remains under investigation.

\section{Noise Sensing Protocol}
\label{app:protocol}

The sensing protocol consists of three stages: initialization of a single photon in the cavity, a sequence of mid-circuit measurements during a variable evolution time, and a final photon population measurement that also serves as a cavity reset.

The experiment begins with the preparation of a single-photon Fock state in the cavity. Starting from the qubit ground state $\ket{g}$, two consecutive $\pi$-pulses are applied to excite the qubit through the ladder $\ket{g}\rightarrow\ket{e}\rightarrow\ket{f}$. A qubit--cavity sideband drive then transfers the excitation to the cavity mode via the transition $\ket{f,0} \rightarrow \ket{g,1}$~\cite{pechal2014microwave,rosenblum2018cnot}, resonant when the sideband drive frequency satisfies $\omega_{\mathrm{sb}} = \omega_{ge} + \omega_{ef} - \omega_\mathrm{c}$, where $\omega_{ge}$ and $\omega_{ef}$ are the transmon qubit transition frequencies.
To verify successful excitation transfer, the qubit is measured immediately after the sideband drive. Runs in which the qubit is not found in $\ket{g}$ are discarded.

\begin{figure}[h]
 \centering 
 \includegraphics[]{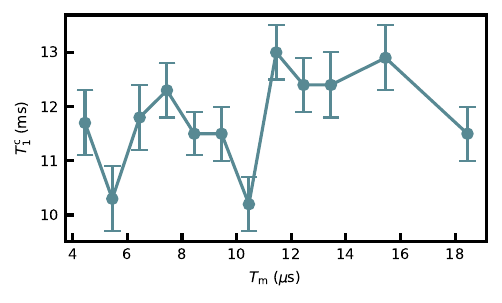}
 \vspace{-2\baselineskip}
 \vspace{10pt} 
 \caption{Intrinsic cavity lifetime as a function of mid-circuit measurement interval duration $T_\mathrm{m}$. The data confirm that the measurement protocol does not degrade the intrinsic single-photon lifetime.
 }
\label{fig:measurement_rate_effect}
\end{figure}

\begin{figure*}
 \centering
 \includegraphics[width=0.9\textwidth]{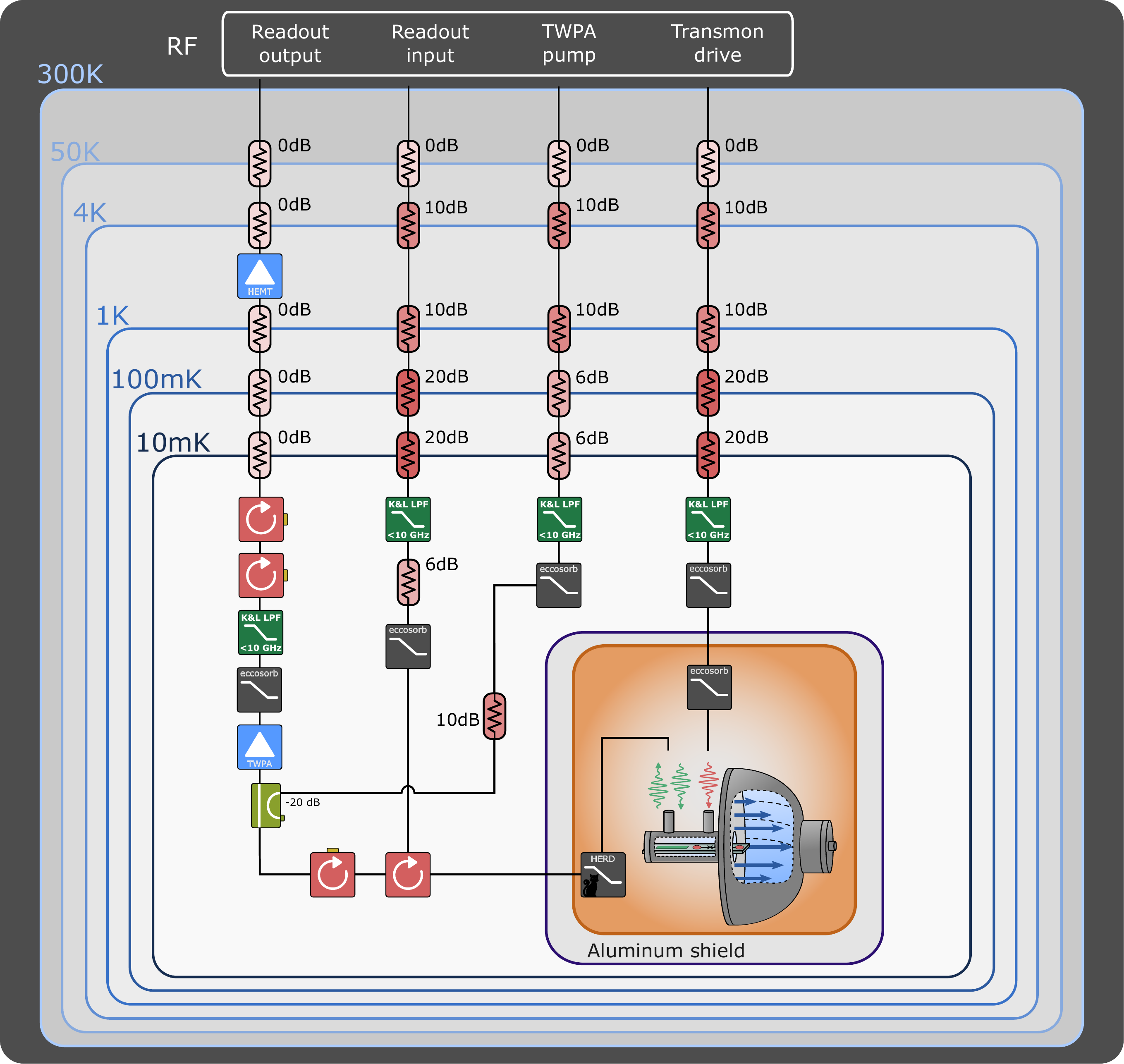}
 \caption{Wiring diagram of the cryogenic microwave setup. The experimental device is placed inside an aluminum shield lined with copper sheets. The control signals are generated using Quantum Machines’ OPX system and upconverted/downconverted using the Octave device. The signals are passed through a series of attenuators, low pass filters (LPF), and Eccosorb infrared filters. The output signal is amplified by a traveling-wave parametric amplifier (TWPA) from Silent Waves before passing through a double-junction isolator, enabling high-fidelity single-shot measurements.}
 \label{fig:fridge_inner_diagram}
\end{figure*}

Following initialization, the system idles for a variable wait time 
during which the qubit is measured every $7.45~\mu\mathrm{s}$, with no measurable effect on the intrinsic single-photon lifetime (Fig. \ref{fig:measurement_rate_effect}). The mid-circuit readout discrimination threshold is set to maximize the post-selection survival fraction, with a false-positive probability of $\sim0.1\%$, and a false-negative probability of $13\%$. The false positive events, together with thermal excitations occurring with a probability of $\bar{n}_{\mathrm{th}}\Gamma T_\mathrm{m} = 0.24\%$ per measurement interval, reduce the fraction of surviving trajectories and hence increase the required number of repetitions.

At the conclusion of the mid-circuit measurement sequence, the cavity
population is measured and reset: a sideband drive transfers any
remaining cavity excitation to the qubit, followed by sequential $\pi_{ef}$ and
$\pi_{ge}$ pulses, so the qubit returns to $\ket{g}$ only if a photon was present. The final qubit state thus encodes the cavity population while the
cavity is reset to vacuum. The final measurement is interpreted as a binary
discrimination between $\ket{g}$ and not-$\ket{g}$, with the decoding of all
four cavity--qubit states summarized in Table~\ref{tab:decoding}.

The only misassigned state is $\ket{e,1}$, in which the photon survives but
the qubit is thermally excited.
Such events slightly reduce the readout contrast but do not bias the extracted rates.

\noindent\begin{minipage}
{\columnwidth}
 {\centering
 \vspace{10pt}\begin{tabular}{l c c c c}
  \toprule
  Initial & After & After $\pi_{ef}$, & Readout & Photon \\
  $\ket{q,n}$ & sideband & $\pi_{ge}$ & outcome & present? \\
  \midrule
  $\ket{g,1}$ & $\ket{f,0}$ & $\ket{g,0}$ & $\ket{g}$      & yes \\
  $\ket{g,0}$ & $\ket{g,0}$ & $\ket{e,0}$ & not-$\ket{g}$ & no \\
  $\ket{e,0}$ & $\ket{e,0}$ & $\ket{f,0}$ & not-$\ket{g}$ & no \\
  $\ket{e,1}$ & $\ket{e,1}$ & $\ket{f,1}$ & not-$\ket{g}$ & \textbf{yes} \\
  \bottomrule
 \end{tabular}
 \par}

 \vspace{6pt}
 \refstepcounter{table}\label{tab:decoding}
 {\footnotesize TABLE~\thetable. Decoding of the final qubit measurement.
 The sideband drive $\ket{f,0}\leftrightarrow\ket{g,1}$ maps a remaining
 cavity photon onto the qubit. The subsequent $\pi_{ef}$ and $\pi_{ge}$
 pulses return the qubit to $\ket{g}$ only in that case. Starting from
 $\ket{e,1}$ the sideband drive is detuned by the transmon anharmonicity and
 the dispersive shift and has no effect. Since the readout discriminates
 $\ket{g}$ from not-$\ket{g}$, a surviving photon is correctly identified in
 all cases except the last row (bold), where a thermal qubit excitation masks
 the surviving photon.\par}
\end{minipage}

\section{Derivation of the post-selected photon survival probability}
\label{app:post-selection_derivation}

We derive the conditional photon survival probability $P_{\ket{1}}^g(t)$ (Eq.~(\ref{eq2})) and the detection probability $\xi$ (Eq.~(\ref{eq3})) for the mid-circuit measurement protocol.

Consider a sequence of $N$ measurement intervals of duration $T_\mathrm{m}$, with total evolution time $t = NT_\mathrm{m}$. In each interval, the cavity photon decays at a total rate $\kappa = \kappa_\mathrm{dd} + \kappa_0$. If the photon is lost via dressed dephasing, the qubit is excited to $|e\rangle$ and subsequently relaxes at rate $\Gamma$.

We first compute the detection probability $\xi$, defined as the probability that a photon lost during a single interval produces a detectable qubit excitation at the time of measurement. Given that the photon is lost during the interval, the probability density for the loss to occur at time $\tau$ after the interval begins is $\kappa e^{-\kappa\tau}/(1 - e^{-\kappa T_\mathrm{m}})$, and the loss is due to dressed dephasing with probability $\kappa_\mathrm{dd}/\kappa$. The qubit must then remain excited for a duration $T_\mathrm{m} - \tau$, which occurs with probability $e^{-\Gamma(T_\mathrm{m}-\tau)}$. Integrating over all possible loss times:
\begin{align}
\xi &= \int_0^{T_\mathrm{m}} \frac{\kappa\, e^{-\kappa\tau}}{1 - e^{-\kappa T_\mathrm{m}}} \cdot \frac{\kappa_\mathrm{dd}}{\kappa} \cdot e^{-\Gamma(T_\mathrm{m} - \tau)}\, d\tau \nonumber \\[4pt]
&= \frac{\kappa_\mathrm{dd}}{1 - e^{-\kappa T_\mathrm{m}}}\, e^{-\Gamma T_\mathrm{m}} \int_0^{T_\mathrm{m}}e^{(\Gamma - \kappa)\tau}\, d\tau \nonumber \\[4pt]
&= \frac{\kappa_\mathrm{dd}}{\Gamma - \kappa} \cdot \frac{e^{-\kappa T_\mathrm{m}} - e^{-\Gamma T_\mathrm{m}}}{1 - e^{-\kappa T_\mathrm{m}}},
\end{align}
yielding Eq.~(\ref{eq3}). In the limit $T_\mathrm{m} \ll 1/\Gamma$, qubit relaxation is negligible and $\xi \to \kappa_\mathrm{dd}/\kappa$, so nearly every dressed-dephasing event is detected. In the opposite limit $T_\mathrm{m} \gg 1/\Gamma$, the qubit relaxes before measurement and $\xi \to 0$. 
False negative errors are neglected in this derivation: since $T_\mathrm{m} \ll 1/\Gamma$, an excitation missed by one measurement is likely to persist to the next, so the probability of eventual detection remains close to unity even for a substantial single-shot false-negative rate.

We now derive $P_{\ket{1}}^g(t)$. Post-selection retains only trajectories in which every mid-circuit measurement returns $|g\rangle$. Within each interval, three outcomes are possible: (i)~the photon survives, with probability $e^{-\kappa T_\mathrm{m}}$; (ii)~the photon is lost and detected (qubit in $|e\rangle$), with probability $(1 - e^{-\kappa T_\mathrm{m}})\,\xi$, which is discarded; or (iii)~the photon is lost but not detected, with probability $(1 - e^{-\kappa T_\mathrm{m}})(1 - \xi)$. Once the photon is lost, no further dressed-dephasing excitations occur, and all subsequent measurements return $|g\rangle$. The total probability of obtaining an all-$|g\rangle$ record is therefore
\begin{align}
P(\text{all } |g\rangle) &= e^{-\kappa t} + (1-\xi)\sum_{j=1}^{N} e^{-\kappa(j-1)T_\mathrm{m}}(1 - e^{-\kappa T_\mathrm{m}}) \nonumber \\
&= e^{-\kappa t} + (1 - \xi)(1 - e^{-\kappa t}),
\end{align}
where the first term corresponds to photon survival and the sum accounts for undetected loss in each interval. The conditional photon survival probability then yields Eq.~(\ref{eq2}):
\begin{equation}
P_{\ket{1}}^g(t) = \frac{e^{-\kappa t}}{e^{-\kappa t} + (1-\xi)(1-e^{-\kappa t})}.
\end{equation}

\section{Simulation of dressed-dephasing dynamics}
\label{app:simulations}

We simulate the joint cavity--qubit dynamics using the
Lindblad master equation formalism, implemented with
QuTiP~\cite{johansson2012qutip}. The Hilbert space is truncated
to at most one cavity photon and two transmon qubit levels. The system
is initialized with one cavity photon and a qubit thermal
population~$n_{\mathrm{th}}$:
\begin{equation}
\rho(0) =
\bigl[
(1 - n_{\mathrm{th}})\, |g\rangle\langle g|
+ n_{\mathrm{th}}\, |e\rangle\langle e|
\bigr]
\otimes |1\rangle_\mathrm{c}\!\langle 1| .
\end{equation}
Coherent noise tones at the cavity--qubit detuning would drive
a coherent excitation exchange between the two modes. Because the
phase of one tone is randomized every $T = 300\;\mathrm{ns}$
(\cref{app:noise_engineering}), much shorter than the system
timescales, this exchange is rendered incoherent and is captured
by dissipative collapse operators rather than a Hamiltonian term.
The collapse operators are:
cavity decay $\sqrt{\kappa_0}\,\hat c$,
transmon relaxation $\sqrt{\Gamma}\,\hat \sigma_-$,
thermal excitation $\sqrt{\Gamma n_{\mathrm{th}}}\,\hat \sigma_+$,
pure dephasing $\sqrt{\Gamma_\phi/2}\,\hat \sigma_z$,
and dressed dephasing
$\sqrt{\kappa_{\mathrm{dd}}}\,\hat \sigma_+\hat c\,$ (forward) and
$\sqrt{\kappa_{\mathrm{dd}}}\,\hat \sigma_-\hat c^\dagger$ (reverse).
The forward and reverse dressed-dephasing rates are equal by the
symmetry of the injected noise spectrum; in practice the reverse process
has a negligible effect since $\kappa_{\mathrm{dd}}/\Gamma \ll 1$.

We compute the transmon excited-state population
$P_e(t) = \mathrm{Tr}[\ket{e}\!\bra{e}\,\rho(t)]$, with all
parameters fixed from independent measurements
(Table~\ref{tab:system_parameters_full}) except
$\kappa_{\mathrm{dd}}$, which is obtained by fitting the
simulation to the measured $P_{\ket{e}}(t)$ (Fig.~\ref{fig2}b).

\section{Bayesian Estimation Methodology}
\label{app:bayesian_statistics}

We extract $\kappa_{\mathrm{dd}}$ and $\kappa_0$ by fitting the
unconditional and post-selected decay curves simultaneously using
Bayesian inference. The unconditional photon survival probability
is modeled as $P_{\ket{1}}(t) = e^{-\kappa t}$ with
$\kappa = \kappa_{\mathrm{dd}} + \kappa_0$, and the post-selected
probability follows Eq. (\ref{eq2}), with $\xi$ computed from
Eq. (\ref{eq3}). At each time point, the number of surviving photons
is modeled as a binomial draw with $n$ total trials and success
probability given by the corresponding model prediction. We assign
uniform priors to $\kappa_{\mathrm{dd}}$ and $\kappa_0$, and a
normal prior to $\Gamma$ centered on an independent $T_1$
measurement with its measured uncertainty.

Posterior samples are obtained using the No-U-Turn Sampler (NUTS)
with 3{,}000 draws and 2{,}000 tuning steps, implemented with
PyMC~\cite{pymc2023}. All reported uncertainties are 68\% highest
density intervals. Posterior predictive checks confirm that the model adequately
describes the data. At high injected noise amplitudes, the
posterior for $\kappa_{\mathrm{dd}}$ is concentrated away from
zero, yielding a well-resolved estimate. At low amplitudes, the
posterior remains consistent with $\kappa_{\mathrm{dd}} = 0$,
yielding only an upper bound and defining the detection floor
visible in Fig.~\ref{fig4}.

\section{Extraction of the cavity-qubit coupling strength}
\label{app:g_extraction}

\begin{figure}[H]
 \centering 
 \includegraphics[]{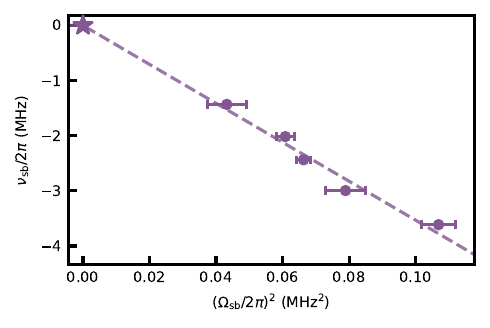}
 \vspace{-2\baselineskip}
 \vspace{10pt} 
 \caption{ The Stark shift of the sideband drive frequency $\nu_\mathrm{sb}$ versus the squared sideband coupling rate $(\Omega_\mathrm{sb}/2\pi)^2$. Error bars are $1\sigma$ bootstrap uncertainties and the dashed line is a weighted fit. The star marks the zero-drive intercept.}
\label{fig:coupling_cal}
\end{figure}

We determine the cavity--qubit coupling $g$ from the parametric sideband transition, driven at
$\omega_\mathrm{sb}=\omega_{ge}+\omega_{ef}-\omega_\mathrm{c} = 2\pi \times3.168~\rm{GHz}$. The drive couples $\ket{f,0}$ to $\ket{g,1}$ at a rate~\cite{pechal2014microwave}
\begin{equation}
\Omega_\textrm{sb}=\sqrt{2}{\beta_\mathrm{sb}}\frac{g}{\Delta}K_\mathrm{q},
\label{eq:gtilde}
\end{equation}
where $\Delta=\omega_\mathrm{c}-\omega_\mathrm{q}$ is the detuning of the main text,
$K_\mathrm{q}=\omega_{ge}-\omega_{ef}>0$ the transmon anharmonicity of
Eq.~(\ref{eq:System_Hamiltonian}), and $\beta_\mathrm{sb}$ the displacement of the qubit induced by the sideband drive. The same displacement Stark shifts the sideband transition frequency $\omega_\textrm{sb}$ relative to its undriven value by $\nu_\mathrm{sb}=-2\eta K_\mathrm{q}|\beta_\mathrm{sb}|^{2}$, where $\eta=\frac{(2\Delta+K_\mathrm{q})(\Delta+K_\mathrm{q})}{\Delta(\Delta-K_\mathrm{q})}\simeq 3.7$ accounts for the stronger Stark shift of $\ket{f}$ with respect to $\ket{e}$ (cf. Eq.~(\ref{eq:stark}) ) -- a factor in agreement with experiment. Eliminating $\beta_\mathrm{sb}$ between the two effects yields 
\begin{equation}
\nu_\textrm{sb} =-\frac{\eta\Delta^2}{K_\mathrm{q} g^2}\Omega_\textrm{sb}^2.
\end{equation}
We measure the sideband transition frequency for increasing drive amplitudes, and plot it against the squared sideband coupling rate extracted from the resulting Rabi oscillations with rate $2\Omega_\textrm{sb}$. We fit the resulting line (Fig.~\ref{fig:coupling_cal}) to yield $g=\Delta\sqrt{\eta/K_\mathrm{q}|m|}=2\pi\times14.8\pm1.0~\mathrm{MHz}$, with $2\pi m=-35~\rm{MHz}^{-1}$ the fitted slope.

\clearpage

\end{document}